\documentclass[useAMS,usenatbib,usegraphicx]{mn2e}
\usepackage{amsmath}
\usepackage[toc,page]{appendix}
\newcommand{\hii}    {H\,{\sc{ii}}}
\newcommand{\mkm}    {$\mu$m}
\newcommand{\tgas}    {$T_{\rm gas}$}
\newcommand{\tdust}    {$T_{\rm dust}$}
\newcommand{\nh}    {$n{\rm (H+H_2)}$}
\newcommand{\kms}    {km s$^{-1}$}
\title[Dust dynamics and evolution in expanding \hii\ regions]{Dust dynamics and evolution in expanding \hii\ regions. \newline I. Radiative drift of neutral and charged grains}
\author[V.V. Akimkin, M.S. Kirsanova, Ya.N. Pavlyuchenkov, D.S. Wiebe]{V.V. Akimkin$^{1}$\thanks{E-mail: akimkin@inasan.ru}, M.S. Kirsanova$^{1}$, Ya.N. Pavlyuchenkov$^{1}$, D.S. Wiebe$^{1}$
\\
$^{1}$Institute of Astronomy of the Russian Academy of Sciences, 48 Pyatnitskaya St. 119017, Moscow, Russia
}
\begin{document}

%\date{Accepted . Received ; in original form \today}

\pagerange{\pageref{firstpage}--\pageref{lastpage}} \pubyear{2015}

\maketitle

\label{firstpage}

\begin{abstract}

We consider dust drift under the influence of stellar radiation pressure during the pressure-driven expansion of an \hii\ region using the chemo-dynamical model MARION.  Dust size distribution is represented by four dust types: conventional polycyclic aromatic hydrocarbons (PAHs), very small grains (VSGs), big grains (BGs) and also intermediate-sized grains (ISGs), which are larger than VSGs and smaller than BGs. The dust is assumed to move at terminal velocity determined locally from the balance between the radiation pressure and gas drag. As Coulomb drag is an important contribution to the overall gas drag, we evaluate a grain charge evolution within the \hii\ region for each dust type. BGs are effectively swept out of the \hii\ region. The spatial distribution of ISGs within the \hii\ region has a double peak structure, with a smaller inner peak and a higher outer peak. PAHs and VSGs are mostly coupled to the gas.  The mean charge of PAHs is close to zero, so they can become neutral from time to time because of 
charge fluctuations. These periods of neutrality occur often enough to cause the removal of PAHs from the very interior of the \hii\ region. For VSGs, the effect of charge fluctuations is less pronounced but still significant. We conclude that accounting for charge dispersion is necessary to describe the dynamics of small grains. 

\end{abstract}


%%%%%%%%%%%%%%%%%%%%%%%%%%%%%%%%%%%%%%%%%%%%%%%%%%%%%%%%%%%%%%%%%
\section{Introduction}

Theoretical and observational studies of dust evolution and survival in \hii\ regions have a long history. The first evidence for dust within an \hii\ region was found from observations of scattered light in the Orion Nebula~\citep{odell_orion} and several other \hii\ regions~\citep{odell_others}. These first observations hinted that dust content in \hii\ regions decreases towards ionizing stars. Development of infrared (IR) detectors has allowed the observation of not only light scattering by dust in \hii\ regions but also proper dust emission \citep[e.g.,][]{na69,sg69,hl71}. These observations have generally confirmed the presence of inner cavities with low dust density in compact and extended \hii\ regions~\citep[e.g.,][]{Gillett1975,aitken,Nakano1983,chini_87,2001A&A...376.1040A}. On the other hand, evidence has also been presented in favour of dust {\it overabundance} inside \hii\ regions \citep{panagia,tt74}.

The interest in the dust in \hii\ regions was renewed by observations of {\it Spitzer Space Telescope}\footnote{{\it Spitzer Space Telescope} is operated by the Jet Propulsion Laboratory, California Institute of Technology, for the National Aeronautics and Space Administration.}, {\it Herschel Space Observatory}\footnote{{\it Herschel} is an ESA space observatory with science instruments provided by European-led Principal Investigator consortia and with important participation from NASA.} and WISE. One of the great {\it Spitzer} results was a discovery of numerous IR emission bubbles in our Galaxy. \cite{churchwell_06} argued that these bubbles are primarily formed by hot young stars in massive star forming regions. Detailed study of the bubbles observed by {\it Spitzer} showed that 86~per~cent of these objects indeed enclose \hii\ regions \citep{deharveng_10}.

The IR bubbles look like partial or closed rings on 8~\mkm\ maps~\citep{churchwell_06,simpson_12}. A similar ring-like morphology is seen in longer wavelength images obtained with {\it Herschel}~\citep{Anderson_12}, in the sense that in the far-IR the bubbles look like rings coinciding with 8~\mkm\ emission rings or residing somewhat further away from the ionizing source \citep{paladini_12}. An IR bubble morphology at 24~\mkm\ is significantly different. Emission at this wavelength is observed not only towards the outer ring but also towards the interior of an \hii\ region \citep{watson_08,deharveng_10}. It must be noted that such a morphology was actually predicted by \cite{wright}.

The different appearance of \hii\ regions at different IR wavelengths must be somehow related to various factors (stellar wind, magnetic field, ultraviolet emission), which leave their imprints on dust properties in the vicinity of hot massive stars. To decipher this information, a detailed model is needed that takes into account not only relevant physical processes, but also a multitude of dust properties. Emission at 8~\mkm\ is generally assumed to belong to PAH molecules excited by ultraviolet (UV) photons, while emission at 24~\mkm\ is presumably generated either by stochastically heated very small grains (VSGs, $a\approx20-50$\,\AA) or by hot sub-micron grains (big grains, BGs). It is natural to attempt to relate the different spatial distributions of emission at near-, mid-, and far-IR wavelengths with spatial distributions of grains having various sizes and/or chemical compositions.

While numerous models of dusty \hii\ regions, both static and dynamic, are presented in the literature, they often either assume that dust is dynamically coupled to the gas or consider dust motion relative to the gas for a single dust component, being primarily directed towards estimating the ability of the dust to pass momentum to the gas \citep[e.g.,][]{KrumholzMatzner2009} or to absorb ionizing radiation within the region \citep[e.g.,][]{1972ApJ...177L..69P,Arthuretal2004}. The first hydrodynamic model of this kind was presented by \cite{mathews_67}, who showed numerically that a central `hole' in the Rosette Nebula can be explained by the radiation pressure on the dust, provided the dust is able to survive in the interior of an \hii\ region. He noted the importance of taking the grain charge into consideration, as a charged grain is more tightly bound to the gas due to Coulomb drag and is also protected against sputtering. Dust was assumed to be dynamically coupled to the gas.

The relative motion of gas and dust under the influence of radiation pressure was considered by \cite{gail_79}. These authors studied the formation of inner cavities via the expansion of dusty cocoons surrounding embedded young massive stars, with the radiation pressure being a main driver for this process. Their model allowed for the gas--dust drift, treated as a diffusion process. \cite{gail_79} also stressed the importance of grain charging, pointing out that the relative gas--dust motion is fastest in regions where the absolute dust charge is small, so that Coulomb drag is less effective. Variations of dust charge along the \hii\ region radius may lead to dust piling up at a certain range of distances from an ionizing star (or stars). As charging properties depend on the grain size, density enhancements for big grains and small grains were shown to occur in different regions.

A detailed analysis of radiation pressure in a static dusty \hii\ region was presented by \cite{Draine2011}, who constructed a family of solutions for various \hii\ region properties and estimated parameters of an inner cavity produced by radiation pressure when the dust is perfectly coupled to the gas. He also considered dust drift and concluded that dust grains can attain quite significant drift velocities (of the order of 100 km s$^{-1}$ and greater) in the inner part of the region, but this drift only affects a minor fraction of the entire dust content of the region. No distinction between various dust species was made in that study.

Spatially resolved IR observations of \hii\ regions draw a complicated picture that apparently cannot be explained by the action of a single dominant mechanism on some generalized dust population. An ideal model would include gas and dust motion, radiation pressure, stellar winds, grain charging, multiple dust populations, etc. We initiated our study of dust evolution in \hii\ regions in \cite{paper_i}. It was shown that images of the RCW120 bubble at 8, 24 and 100~\mkm\ can be reproduced simultaneously if PAHs are partially destroyed inside the \hii\ region, but the abundance of VSGs and BGs stays close to its mean interstellar value. Dust emission at 24~\mkm\ in that model is mainly produced by stochastically heated VSGs as in the general interstellar medium \citep{DIMmodel}, while 100 \mkm\ emission is generated by cold BGs in the dense shell surrounding the \hii\ region.

The model from \cite{paper_i} is based on the simplifying assumption that the dust is dynamically coupled to the gas and the only process that may alter the dust properties is the destruction of PAHs. In the present work, we study another, less explored aspect of dust evolution in an \hii\ region. Namely, we neglect dust destruction for the time being, but relax an assumption of tight coupling between the dust and gas and study how the differential dust drift under the action of radiation pressure and {\em grain charging\/} influence the spatial distribution of dust grains of different sizes and types within an \hii\ region.

%%%%%%%%%%%%%%%%%%%%%%%%%%%%%%%%%%%%%%%%%%%%%%%%%%%%%%%%%%%%%%%%%
\section{Modelling}\label{sec_model}

The presented model is based on the chemo-dynamical code MARION, which is being developed as an extension of the Zeus-2D code \citep{stone_92}. The MARION code is designed to simulate a pressure-driven expansion of an \hii\ region around a young massive star. It is assumed that a star is embedded in a molecular cloud with a uniform gas distribution, characterized by number density $n$. At time $t=0$, the star starts ionizing and heating the surrounding gas. The size of the \hii\ region grows because of the gas pressure difference between hot ionized and cold molecular gas \citep[e.g.][]{mathews_65,spitzerbook}. A shock front develops ahead of the ionization front and propagates into the molecular cloud, being preceded by fronts of H$_2$ and CO dissociation \citep[see e.g.][]{tielens_85,hosokawa_06,paper_hiimodel}. This shock front shovels molecular gas into a dense envelope around the \hii\ region.

The MARION code is described in detail in \cite{paper_hiimodel} and \cite{paper_i}. Here we recapitulate its main features. The expansion of an \hii\ region is modelled in a one dimensional approximation, taking into account a detailed thermal balance of the gas and dust along with the related chemistry. In this work we do not consider the chemical evolution of the molecular gas, but we still have to account for the simple chemistry of H$_2$, CO, H$_2$O and OH molecules, which is needed to model heating and cooling processes in a photon dominated region. Cooling in lines of neutral and ionized O, S and Si, which is important for the thermal balance of an \hii\ region, is also included. A summary of the thermal processes is given in Table~\ref{heatandcool}.

Photoreaction rates are calculated by integration of corresponding cross-sections, taken from the Leiden database~\citep{vdh_88,faraday_06}, with the local emission spectrum used at each point of a computation grid. Parameters of other reactions are taken from the UMIST~95 database~\citep{umist_95}. Initial physical parameters and non-zero abundances of chemical species are listed in Table~\ref{init_model}. A full list of chemical species includes H, H$^+$, H$_2$, CO, C, C$^+$, O, O$^+$, O$^{++}$, S, S$^+$, S$^{++}$, Si, Si$^+$, OH, H$_2$O and He. Surface species are not taken into account to save computational time. Initial abundances in Table~\ref{init_model} are based on the interstellar medium-like set of abundances from the CLOUDY model~\citep{cloudy_13}. The spectrum of ionizing stellar emission is taken from~\citet{kurucz_79}.

For this work, we further upgraded the MARION code and added a treatment for charged grain motion relative to the gas. Two primary forces acting on a dust grain are radiation pressure and aerodynamic drag. The radiation pressure force is given by
\begin{equation}
F^g_{\rm rp}=\frac{1}{c}\int C^g_{\rm rp}(\nu) F(\nu)\,d\nu,
\end{equation}
where $c$~is the speed of light, $F(\nu)$ is the radiation flux computed in the model of the \hii\ region, $C^g_{\rm rp}(\nu)$ is the radiation pressure cross-section for the grain type $g$:
\begin{equation}
C^g_{\rm rp}(\nu)=C^g_{\rm abs}(\nu)+(1-\langle\cos\theta^g\rangle)C^g_{\rm sca}(\nu).
\end{equation}
Absorption and scattering cross-sections for carbonaceous and silicate dust grains, $C^g_{\rm abs}$ and $C^g_{\rm sca}$, are calculated using Mie theory. PAH absorption cross-sections are adopted from~\cite{2007ApJ...657..810D}.

The grain charge is determined by the balance between electron and ion accretion on to the grain and photoelectric emission~\citep{tielensbook} (see Appendix~\ref{AppA} for details). To save computational time, the grain charge is pre-calculated on a grid of electron number densities, gas temperatures and distances from the star (or radiation field strength).

A grain of radius $a_{\rm gr}$ and charge $Z_{\rm gr}$, moving with velocity $v$ relative to the gas, undergoes braking due to collisions with gas particles. Neglecting surface processes, the resulting aerodynamic force is~\citep{DraineSalpeter1979}:
\begin{align}
  F_{\rm drag} &= 2\pi a_{\rm gr}^2 k T_{\rm gas} \left\{\sum\limits_i n_i \left[\vphantom{z^2}G_0(s_i)+\right.\right. \nonumber \\
   &+ \left.\left. z_i^2\phi^2\ln(\Lambda/z_i)G_2(s_i)\right]\vphantom{\sum\limits_i}\right\}, \label{lFdrag}
\end{align}
where
\begin{align}
s_i\equiv\sqrt{m_iv^2/(2kT_{\rm gas})},\nonumber\\
G_0(s_i)\approx8s_i/(3\sqrt{\pi})\sqrt{1+9\pi s_i^2/64},\nonumber\\
G_2(s_i)\approx s_i/(3\sqrt{\pi}/4+s_i^3),\\
\phi\equiv Z_{\rm gr}e^2/(a_{\rm gr}\times kT_{\rm gas}),\nonumber\\
\Lambda\equiv3/(2a_{\rm gr}e|\phi|)\sqrt{kT_{\rm gas}/(\pi n_e)}.\nonumber
\end{align}
Here $T_{\rm gas}$ is the gas temperature, $e$~is an elementary charge, $k$~is the Boltzmann constant, $n_e$~is the electron number density. Summation over $i$ involves number densities $n_i$ and masses $m_i$ of all neutral and charged gas components. We consider H, H$_2$, H$^+$, He and electron number densities. Their radial profiles are taken from the results of chemo-dynamical simulation at a considered time moment.

The gas velocity $v_{\rm gas}$ is given by the hydrodynamic part of the MARION code. We assume that a grain always moves with the terminal velocity $v_{\rm gr}$, which is given by the grain equation of motion, provided that the grain acceleration is zero. In other words, we assume that as the grain moves through the \hii\ region, its velocity quickly adapts to a local environment. In this approximation
\begin{equation}
 F_{\rm drag} (a_{\rm gr}, Z_{\rm gr}, v=v_{\rm gas}-v_{\rm gr}) + F_{\rm rp}(a_{\rm gr})=0.
\label{dust_dynamics}
\end{equation}
Solving this non-linear equation relative to $v_{\rm gr}$ for grains of different types, one can simulate the motion of various dust grains in the \hii\ region. We have tested this assumption for integrating dust equations of motion in a static \hii\ region pre-computed with the CLOUDY code~\citep{cloudy_13} and found that the time needed to reach the terminal velocity almost never exceeds 200 yr.

\begin{table}
\caption{Heating and cooling processes.}
\label{heatandcool}
\begin{tabular}{@{}ll}
\hline
Process & Reference\\
\hline
\multicolumn{2}{|c|}{\it Heating} \\
Photoionization of H       &\citet{tielensbook}\\
Photoionization of C       &\citet{jonkheid04}\\
Photoelectric heating      &\citet{jonkheid04}\\
H$_2$ formation            &\citet{jonkheid04}\\
H$_2$ dissociation         &\citet{jonkheid04}\\
H$_2$ pumping             &\citet{jonkheid04}\\
\hline
\multicolumn{2}{|c|}{\it Cooling} \\
Recombination of H$^+$        &\citet{tielensbook}\\
Lyman $\alpha$ emission          &\citet{tielensbook}\\
Hydrogen free-free transitions&\citet{tielensbook}\\

[OI]$\lambda=63.2 \mu$m       &\citet{drainebook}\\

[OI]$\lambda=6300.3$\AA       &\citet{tielensbook}\\

[OII]$\lambda=3728.8$\AA      &\citet{hollenbach89}\\

[CII]$\lambda=157.7\mu$m    &\citet{tielensbook}\\

[SII]$\lambda=6730.8$\AA       &\citet{tielensbook}\\

[OIII]$\lambda=5006.9$\AA      &\citet{tielensbook}\\

[OIII]$\lambda=51.8 \mu$m     &\citet{tielensbook}\\

[SIII]$\lambda=9530.9$\AA      &\citet{tielensbook}\\

[SIII]$\lambda=187.0\mu$m    &\citet{tielensbook}\\

[SiII]$\lambda=35.0\mu$m    &\citet{drainebook}\\

CO rot\&vib transitions    &\citet{hollenbach79}\\
H$_2$ rot\&vib transitions    &\citet{neufeld93}\\
H$_2$O rot\&vib transitions   &\citet{hollenbach79}\\
OH rot\&vib transitions    &\citet{hollenbach79}\\
\hline
Dust-gas interaction &\citet{tielensbook}\\
\hline
\end{tabular}
\end{table}    

\begin{table}
\caption{Initial conditions for modelling of an expanding \hii\ region.}
\label{init_model}
\begin{tabular}{@{}lc}
\hline
Quantity & Value\\
\hline
\tgas & 10~K\\
\tdust & From balance equation\\
\nh & $3\times 10^3$~cm$^{-3}$\\
\hline
$x(\rm{S})$ & $3.24\times 10^{-5}$\\
$x(\rm{Si})$ & $3.16\times 10^{-6}$\\
$x(\rm{H_2})$ & 0.5\\
$x(\rm{CO})$ & $2.51\times 10^{-4}$\\
$x(\rm{H_2O})$ & $6.8\times 10^{-5}$\\
$x(\rm{He})$ & $9.8\times 10^{-2}$\\
\hline
\end{tabular}
\end{table}

In the numerical model we define $v_{\rm gas}$ and $v_{\rm gr}$ at cell borders and use them to calculate the corresponding fluxes across borders. Equation~\eqref{dust_dynamics} is solved with the bisection method for every cell. As soon as all $v_{\rm gr}$ are found, we use them to calculate dust fluxes $f_{\rm gr}$ in the same way as in the transport step of the hydrodynamic solver \citep{stone_92}. The calculated fluxes are used to update the dust density according to the equation of mass conservation:
\begin{equation}
\frac{\partial n}{\partial t} + \frac{1}{r^2} \frac{\partial}{\partial r}(r^2 f_{\rm gr}) = 0.
\end{equation}
In this formulation, dust is dynymically coupled to the gas when $v_{\rm gr}=v_{\rm gas}$.

Dust drift is considered on top of the hydrodynamic model, in the sense that the motion of grains is modelled simultaneously with the \hii\ region expansion, but without feedback. Specifically, to model dust drift and to calculate dust densities we use the following approximations. First, we neglect momentum transfer from the dust to the gas caused by collisional drag. This allows us to decouple the equations of dust dynamics from the full system of model equations and to treat them independently. Second, we treat each dust species as an individual component neglecting grain destruction. All the dust particles of a certain type within a cell of the hydrodynamic grid have the same velocity. Third, we assume that drifting dust grains affect neither the gas temperature nor the gas dynamics. To take into account dust absorption of UV radiation in the hydrodynamic and chemical model of an \hii\ region, we use a `generalized' dust which has a mean radius of $1.3\times10^{-5}$~cm and is mixed with gas with a uniform mass fraction of 1~per cent. Thus, our model is inconsistent in the dust treatment, but a much more complicated model would be needed to achieve consistency in this respect.

The motion of charged grains in a magnetic field is also affected by the Lorenz force. In this paper, the interaction of charged grains with a magnetic field is not considered as this would require knowledge of the magnetic field geometry in the \hii\ region. The strength of the magnetic field has been studied observationally by \citet{2011ApJ...736...83H}, while its large-scale morphology has been modelled by a number of authors, including \citet{2007ApJ...671..518K,2012ApJ...745..158G,2011MNRAS.414.1747A,2011MNRAS.412.2079M}. If the large-scale magnetic field is regular, we may expect that dust grains would migrate mostly along the magnetic field lines, which would lead to oblate or arc-like morphologies of dust distribution. So, the radial dust density profiles presented in the next section could be mostly relevant for regular magnetic field geometries with some factor depending on the magnetic field strength and orientation. A magnetic field with a significant irregular component requires much more sophisticated modelling.

\section{Dust drift in an expanding \hii\ region}\label{sec_drift}

In this study we do not attempt to describe in detail any particular object, but nevertheless some fiducial parameters are needed to model an expanding \hii\ region. Thus, we adopted the observed data for the  \hii\ region of RCW~120, which has an almost perfect ring-like shape~\citep{dehaveng_05,deharveng_09,zavagno_10}, so it is a good prototype for our simulation. Its parameters are given in Table~\ref{tab_rcw120}.

\begin{table*}
\caption{Parameters of RCW~120 used in the modelling.}
\label{tab_rcw120}
\begin{tabular}{@{}lll}
\hline
Quantity & Value & Reference\\
\hline
$T_{\rm star}$ & 35000~K &\cite{georgelin_70,crampton_71,martins_2010,diaz_miller}\\
$R_{\rm star}$ & 6$R_{\sun}$ & \cite{diaz_miller}\\
\nh$_{\rm init}$ & $3\times 10^3$~cm$^{-3}$& \cite{zavagno_07}\\
$n({\rm H^+})$ & 86~cm$^{-3}$ &\cite{zavagno_07}\\
$R_{\rm HII region}$ & 4\arcmin\ or 1.5~pc & Adopting distance of 1.3~kpc from \citet{russeil_03}\\
\hline
\end{tabular}
\end{table*}

\begin{figure}
\begin{center}
\includegraphics[scale=0.65,angle=270]{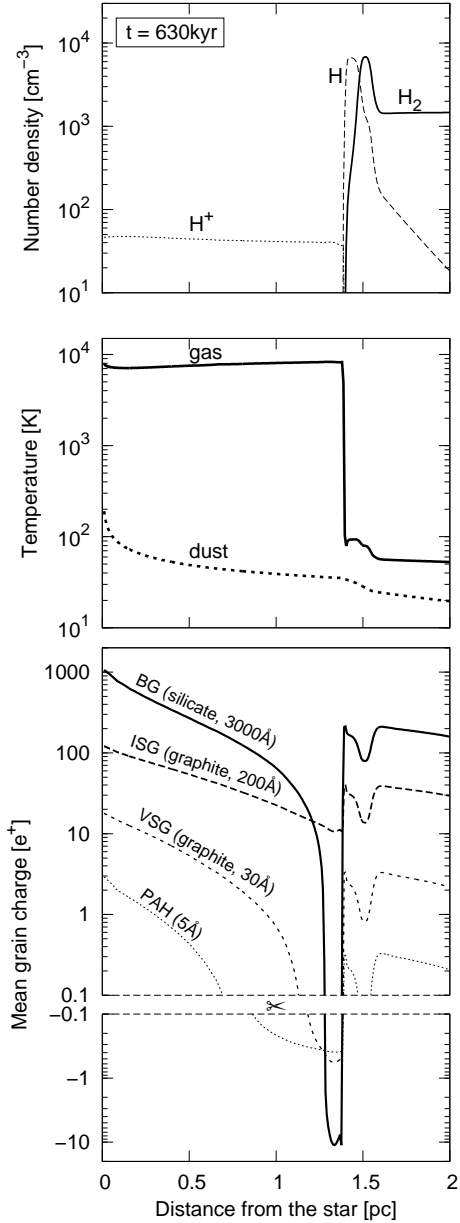}
\caption{Top panel: Number density of ionized and neutral gas components for $t=630$~kyr. Middle panel: Gas and dust temperatures. Bottom panel: Mean charges of PAHs, very small and intermediate graphite grains, and big silicate grains. The bottom panel is split into two parts to present positively and negatively charged grains.}
\label{fig:structure}
\end{center}
\end{figure}

In our model, the configuration that approximately corresponds to RCW\,120 is reached in about 630~kyr after the onset of ionization. The distributions of ionized, atomic and molecular hydrogen at this time are shown in Fig.~\ref{fig:structure} (top panel). 

The hydrogen number density within the \hii\ region is about 40 cm$^{-3}$, half the observational estimate for RCW\,120~(see Table~\ref{tab_rcw120}). The region is surrounded by a dense neutral envelope, which was swept-up by the expanding shock front. The enhanced density in the compressed envelope is about five times the density in the molecular region. The thermal structure of the region is illustrated in the middle panel of Fig.~\ref{fig:structure}.

We present results for four grain sizes, three of which (PAHs, VSGs, and BGs) correspond to `conventional' dust grains that are typically considered in various dust studies. We also consider grains that are bigger than VSGs but smaller than BGs. These intermediate-sized grains (ISGs) are added to the model to estimate possible outcomes of a more detailed treatment of the dust size distribution. The parameters of the dust types studied are summarized in Table~\ref{tab:dustparam}.

\begin{table}
\caption{Adopted dust types.}
\label{tab:dustparam}
\begin{tabular}{lcc}
\hline
    & Composition & Radius (\AA) \\
\hline
Polycyclic aromatic hydrocarbons   & --      & $5$  \\
Very small grains (VSGs)   & graphite & $30$  \\
Intermediate size grains (ISGs) & graphite & $200$  \\
Big grains (BGs)    & silicate & $3000$  \\
\hline
\end{tabular}
\end{table}

The grain charge is determined by the balance between competing processes of electron and ion accretion and photoelectric emission. Illumination conditions inside the \hii\ region mostly drive grains to positive charges (see the bottom panel of Fig.~\ref{fig:structure}). For reference, maximum grain charges for 13.6~eV photon absorption, such that the next photoelectric emission is prevented by the electrostatic barrier, are +3, +19, +128 and +1167\,$e$ for PAHs, VSGs, ISGs and BGs, respectively \citep[Eq.~(22) from~][]{2001ApJS..134..263W}. The further away we move from the star, the less effective photoelectric emission becomes due to radiation absorption and dilution. At the same time, the electron accretion rate is pretty much the same over the entire \hii\ region. This is why the mean charge within the ionized region decreases outwardly and reaches negative values for PAHs, VSGs and BGs. ISGs are somewhat special in this respect as the negative gradient of their charge is 
much shallower than for other grain types, and they do not become negative at all. The reason is that dust grains of this size absorb photons (and emit electrons) very effectively. The negative gradient for grain charges is quite important for the resulting dust density profiles, as will be discussed later. The sharp charge rise at the edge of the ionized region is caused by a sharp decrease in the electron number density, and the dip located further away is related to the enhanced electron accretion rate at the density bump.

As details of dust drift sensitively depend on a grain charge, below we present results for neutral and charged grains separately. Also we will check how important a specific way of taking the grain charge into account is. Possible options are to consider the mean charge only for grains of a given size, the charge distribution for an ensemble of single-size grains or the fluctuating charge for a grain. These approaches produce results with different levels of detail (and computational cost), so it may be desirable to find the simplest necessary approach.

\subsection{Neutral grains}

\begin{figure*}
\includegraphics[scale=0.5,angle=270]{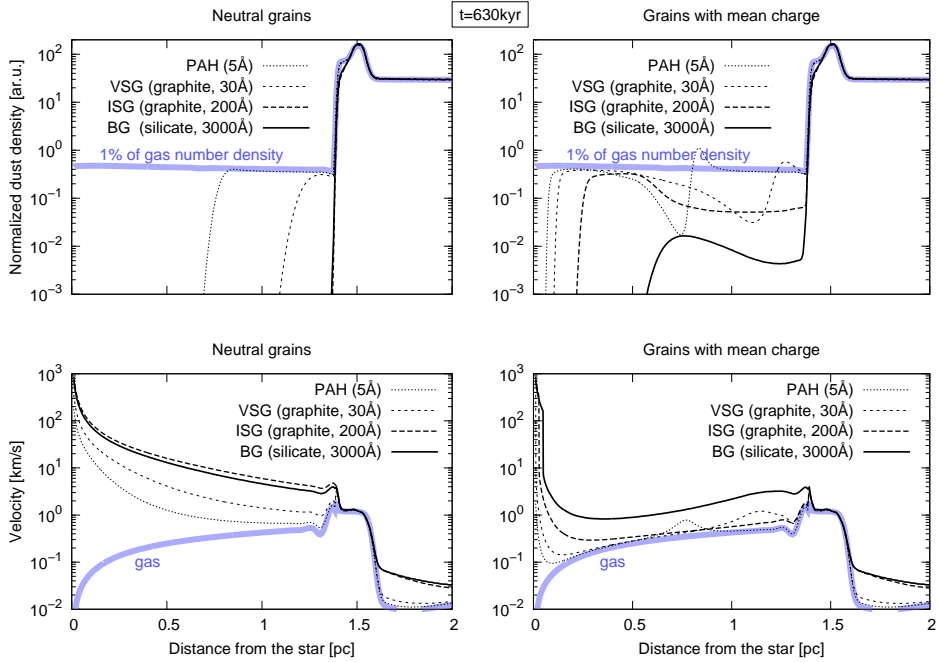}
\caption{Top panels: Density profiles of neutral (left) and charged (right) PAHs, VSGs, ISGs and BGs normalized to 1~per~cent of the gas number density. Bottom panels: Radial velocities of neutral (left) and charged (right) PAHs, VSGs, ISGs and BGs along with gas velocity. The time since the beginning of the expansion is $630$~kyr.}
\label{fig:neutral_charged}
\end{figure*}

Grains within an \hii\ region are definitely not neutral, so we present a calculation of dust drift for neutral PAHs, VSGs, ISGs and BGs only as a test case. It is assumed that initially all the dust components are well mixed with the gas. When the expansion of the \hii\ region starts, radiation pressure pushes grains away from the star, and the only force that counteracts their motion is the collisional drag. As the gas density is not very high, grain velocities for all the four grain types significantly exceed the gas velocity as shown in the bottom left-hand panel of Fig.~\ref{fig:neutral_charged}, being greater than 100~\kms\ in the immediate star vicinity. The difference becomes smaller than 10~\kms\ only near the border of the ionized region.

The bigger the particle, the weaker it is coupled to the gas, so ISGs and BGs are swept out of the \hii\ region completely (their density profiles coincide on the graph), while PAHs retain their initial abundance in the outer part of the region (remember that dust destruction is not taken into account), as seen in the top left-hand panel of Fig.~\ref{fig:neutral_charged}. By 630 kyr, about 25~per cent and 50~per cent of the \hii\ region volume is free from PAHs and VSGs, respectively. The gas and dust velocities are the same in the compressed neutral envelope, where the dust is coupled to the gas. On this and other similar plots, all the dust densities are normalized to 1~per~cent of the gas number density. This normalization emphasizes the difference between density profiles and drag regimes for grains of different sizes. Their actual relative and absolute abundances should be estimated using emission modelling and other constraints set by general dust models. This will be presented in a subsequent study.

\subsection{Grains with mean charge}

The dynamics of charged grains differs drastically from that of neutral grains due to additional Coulomb drag. As a first step in the drift modelling of charged grains, we considered the drift of grains with a representative mean charge. Whether the representative charge leads to a representative density profile will be discussed later. Velocities and spatial distribution of PAHs, VSGs, ISGs and BGs for $t=630$~kyr are shown on the right-hand panels of Fig.~\ref{fig:neutral_charged} (their mean charges are plotted on the bottom panel of~Fig.~\ref{fig:structure}). Obviously, the spatial distributions of charged grains are not the same as those of neutral ones. Coulomb drag notably slows down the grain motion, making the dust more coupled to the gas. The velocities of charged PAHs, VSGs and ISGs only slightly exceed the gas velocity in most of the \hii\ region. This is why charged particles, unlike neutral particles, stay relatively abundant in the ionized gas. On the other hand, dust drift due to radiation pressure has a profound 
effect on both charged and neutral BGs, significantly decreasing their density in the ionized gas in both cases, even though the abundance of charged BGs exceeds significantly the abundance of BGs in the test case of zero charges.

The velocities of all grains become greater than 100~\kms\ within about 0.1~pc from the star. Thus, the vicinity nearest to the star is cleaned from dust completely. Note the discontinuity in the velocity profile of BGs close to the star, which appears because the equation of motion with the Coulomb term may have three solutions for the velocity \citep{Draine2011}. In the general case, it has two stable solutions (low and high velocity) and one unstable solution (intermediate velocity). In most of the \hii\ region, we pick the stable low velocity solution as being more physical, but this solution disappears in the star's vicinity, so the high velocity solution is presented.  

There are two drift regimes depending on whether there is a notable zone where the grains have zero or close to zero charge (PAHs and VSGs). In such regions, the Coulomb drag force vanishes and the grains fly outside faster till they enters the region where their charge becomes negative and where the Coulomb drag force again becomes significant. This leads to a pile-up of PAHs and VSGs at 0.8 and 1.3~pc, respectively. For bigger grains, the zero charge zone is absent (ISGs) or has much smaller extent (BGs), so another drift regime operates. While the radiation pressure force decreases outside, which slows the grain drift, the grain charge also decreases, lowering the grain coupling to the gas and increasing its velocity. For the considered physical parameters, $F_{\rm rp}^{g}/F_{\rm drag}$ increases outside, so the grain velocity also increases outside. As an outer part of the ionized zone is cleared more effectively from grains compared to the inner part, a dust density bump emerges inside the \hii\ region.

So, we see that when we take the mean charge into account, a dust density bump emerges inside the \hii\ region. This is an intriguing finding in view of the double-ring geometries of \hii\ regions in the mid-IR. Let us now check whether these features are preserved when we introduce a more detailed description of grain charging.

\subsection{Grains with charge distribution}

So far we have assumed that all the grains of a certain type have the same charge at a given time and at a given location. However, grains in the ensemble of equal-sized grains will attain slightly different charges, so that there will be a charge distribution with a non-zero dispersion. The dispersion is proportional to the charge value, so bigger grains have larger charge scatter \citep[see eqs. (5.81)--(5.82) in][]{tielensbook}. We use two approaches to take into account the impact of the charge dispersion on the dust density profiles.

\begin{figure*}
\includegraphics[scale=0.5,angle=270]{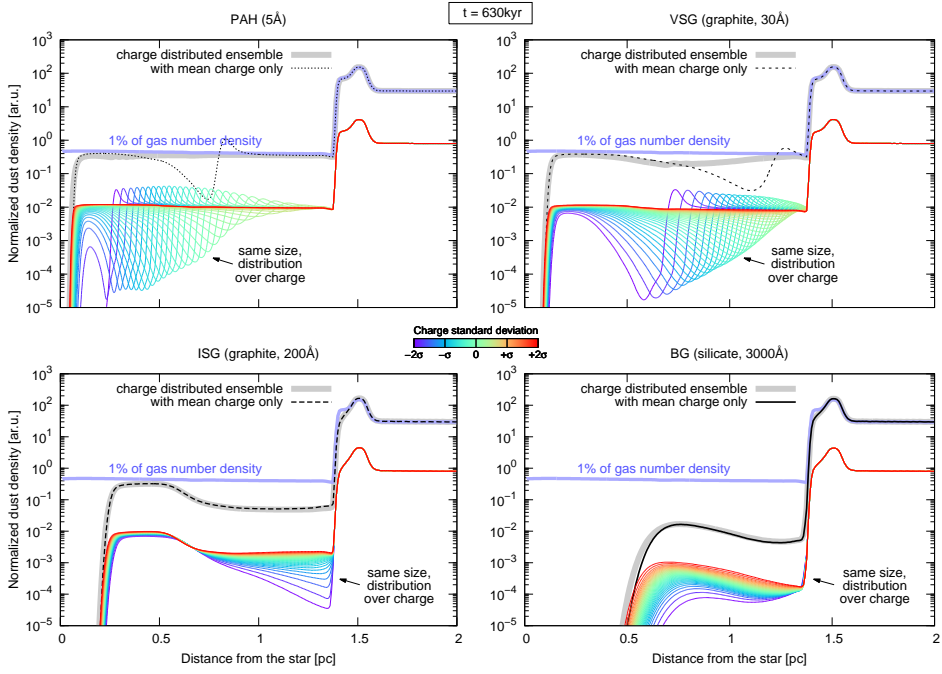}
\caption{Density profiles of charged PAHs, VSGs, ISGs and BGs for $t=630$~kyr normalized to 1~per~cent of the gas number density. Colour lines illustrate density variations due to charge dispersion within an ensemble of single-sized grains.}
\label{fig:37}
\end{figure*}

In the {\em first approach\/}, we consider an ensemble of single-type grains with a distribution of charges, which is characterized by a standard deviation $\sigma_z$ for a given location and time. We consider 37 bins for grain charge, assuming that each bin contains the same fraction of the total dust mass, and trace grain motion for each bin separately. In this approach, the charge of a grain changes with time in accord with the mean charge and standard deviation for this grain type, so that $Z_i=\overline{Z}+f_i\sigma_z$ with factors $f_i\neq f_i(t)$ allowing an even mass distribution between bins. In the first approach, grains are allowed to have a fractional charge as we try to trace the motion of the whole ensemble, rather than of individual grains. Density profiles for individual bins are shown in Fig.~\ref{fig:37} as rainbow-like groups of lines. The sum of individual profiles is shown by the thick grey line. For comparison, we also show the density profiles for grains with the mean charge (black lines).

Density radial profiles for ISGs and BGs are nearly the same as when all the grains have the same (mean) charge. The depression in the ISGs density radial profile on the inner side of the dense shell becomes deeper for grains having a charge smaller than the mean. Similar behaviour is observed for BGs as well. The density in the ionized region is lower for grains with lower charge and is higher for grains with higher charge. But they all sum up to nearly the same total density as when there is no charge dispersion.

On the other hand, the effect of the grain charge distribution for PAHs and VSGs is profound. While density profiles for grains in each charge bin show the same general morphology, with the inner density bumps and the pile-up peaks, the location of the pile-up peak is a strong function of the grain charge. The peak is located much closer to the star for grains at the lower boundary of the considered charge range and is nearly absent for grains at the upper boundary of the range. The net result is that the actual density profile is quite flat, with all the pile-up peaks spread evenly over the ionized region.

\begin{figure*}
\includegraphics[scale=0.55,angle=270]{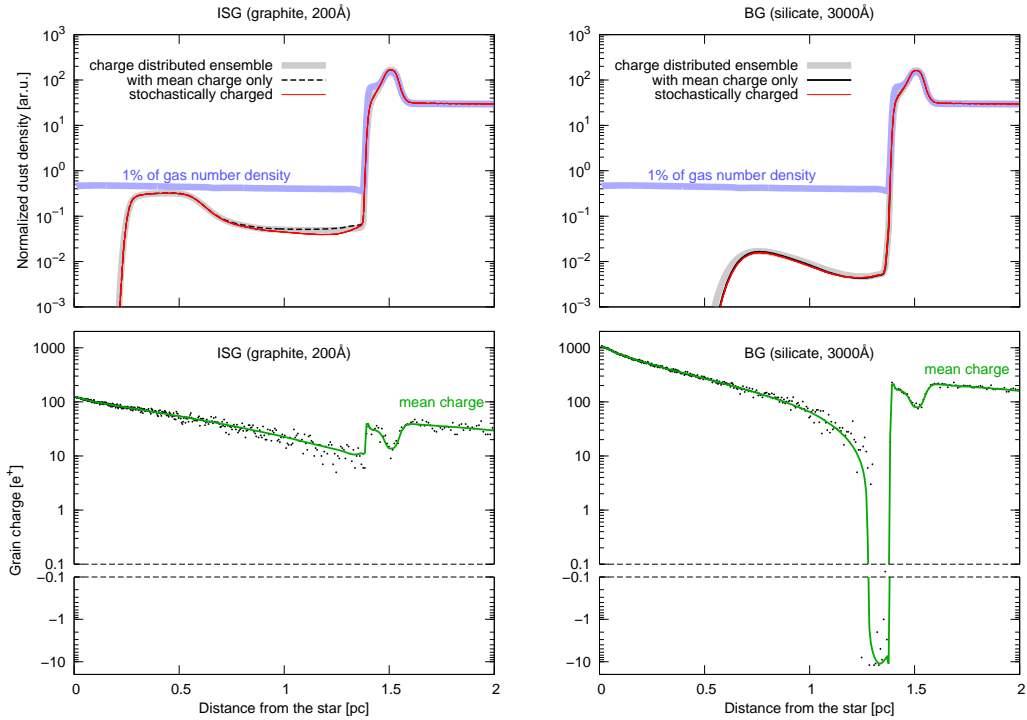}
\caption{Comparison of approaches for grain charge evolution. Top panels: Density profiles for ISGs and BGs normalized to 1~per~cent of the gas number density. Red lines show density profiles for grains with a fluctuating charge. The thick grey line and black lines are the same as in Fig.~\ref{fig:37}.  Bottom panel: ISGs and BGs charges (dots between upper and lower sub-panels represent zero charges).}
\label{fig:stochastic_a}
\end{figure*}

\begin{figure*}
 \includegraphics[scale=0.55,angle=270]{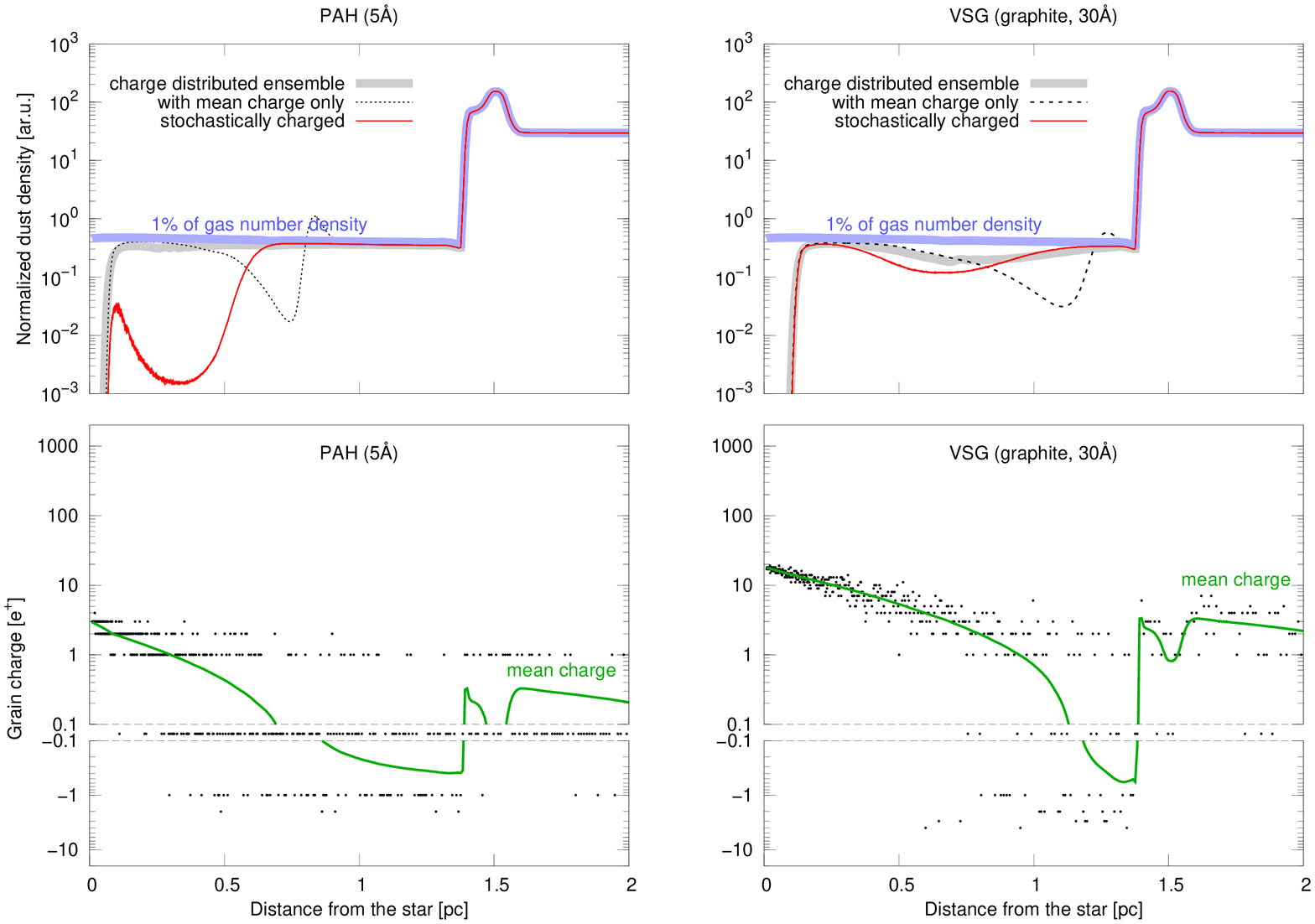}
 \caption{Same as Fig.~\ref{fig:stochastic_a} but for PAHs and VSGs. Fluctuations in the density profile of PAHs (red lines) reflect dispersion due to different random seeds in the stochastic charge modelling. For other grain types, these fluctuations are smaller than the line width.}
 \label{fig:stochastic_b}
\end{figure*}

In the approach, we have just described, we assume that the grain charge changes only in response to changes in the physical environment in the sense that both $\overline{Z}$ and $\sigma_z$ change with time, but $f_i$ stays constant. But the charge distribution arises not only because there are grains with different charges, but also because the charge on each particular grain fluctuates with time. In the {\em second approach\/}, we consider grains with a fluctuating charge, which is set to a Gaussian random value with a given mean and standard deviation at every hydrodynamic time step (only integer charge values are allowed). Five realizations are considered with different random seeds. The results are shown in Fig.~\ref{fig:stochastic_a} for ISGs and BGs and in Fig.~\ref{fig:stochastic_b} for PAHs and VSGs.

Again, the results for ISGs and BGs are nearly the same as when all the grains are assumed to have the mean charge determined by the current physical environment. Their mean charge is large, and its fluctuations are not significant, so that the dust density profiles for the three considered cases do not differ much.

The situation is somewhat different for VSGs. As we mentioned above, a realistic grain charge distribution tends to smooth the resulting density profile in comparison with when all the VSGs are assumed to have the mean charge. This tendency is preserved also when each grain is assumed to have a fluctuating charge. It does not really matter how exactly the charge distribution is taken into account for VSGs. Even though charge excursions from the mean for a single grain can be quite significant (bottom right-hand panel of Fig.~\ref{fig:stochastic_b}), the resulting density profile (top right-hand panel of Fig.~\ref{fig:stochastic_b}, red band) is very similar to the profile for the grain ensemble with a non-fluctuating charge distribution (top right-hand panel of Fig.~\ref{fig:stochastic_b}, grey band).

Finally, PAHs behave differently in all three considered cases. Their mean charge can attain both positive and negative values, ensuring effective coupling between PAHs and the gas over most of the \hii\ region and causing the pile-up peak to appear near the distance range where the charge is zero. In the model with the grain ensemble, positively and negatively charged PAHs are mixed with each other, and all the density profile features are smeared out. The situation changes dramatically when we add charge fluctuations for each particular PAH particle. Their mean charge is quite small, so the charge of a PAH particle fluctuates between --2 and +3, being neutral for a significant portion of the time (bottom left-hand panel of Fig.~\ref{fig:stochastic_b}). As a result, the distance range in which the PAHs are decoupled from the gas is significantly wider than  when all the PAHs are assumed to have the mean charge. An inner cavity is formed due to the effective escape of PAHs from the star's vicinity when PAH particles have zero charge.

We conclude that the mean charge is representative in terms of dust dynamics for larger grains (ISGs and BGs), but a more sophisticated consideration is needed for smaller grains (PAHs and VSGs). For PAHs, a fluctuating grain charge seems to be the only realistic approach among the approaches we have considered. This approach is preferable for VSGs also.

\subsection{Charged grains: chemical composition}

\begin{figure*}
\includegraphics[scale=0.5,angle=270]{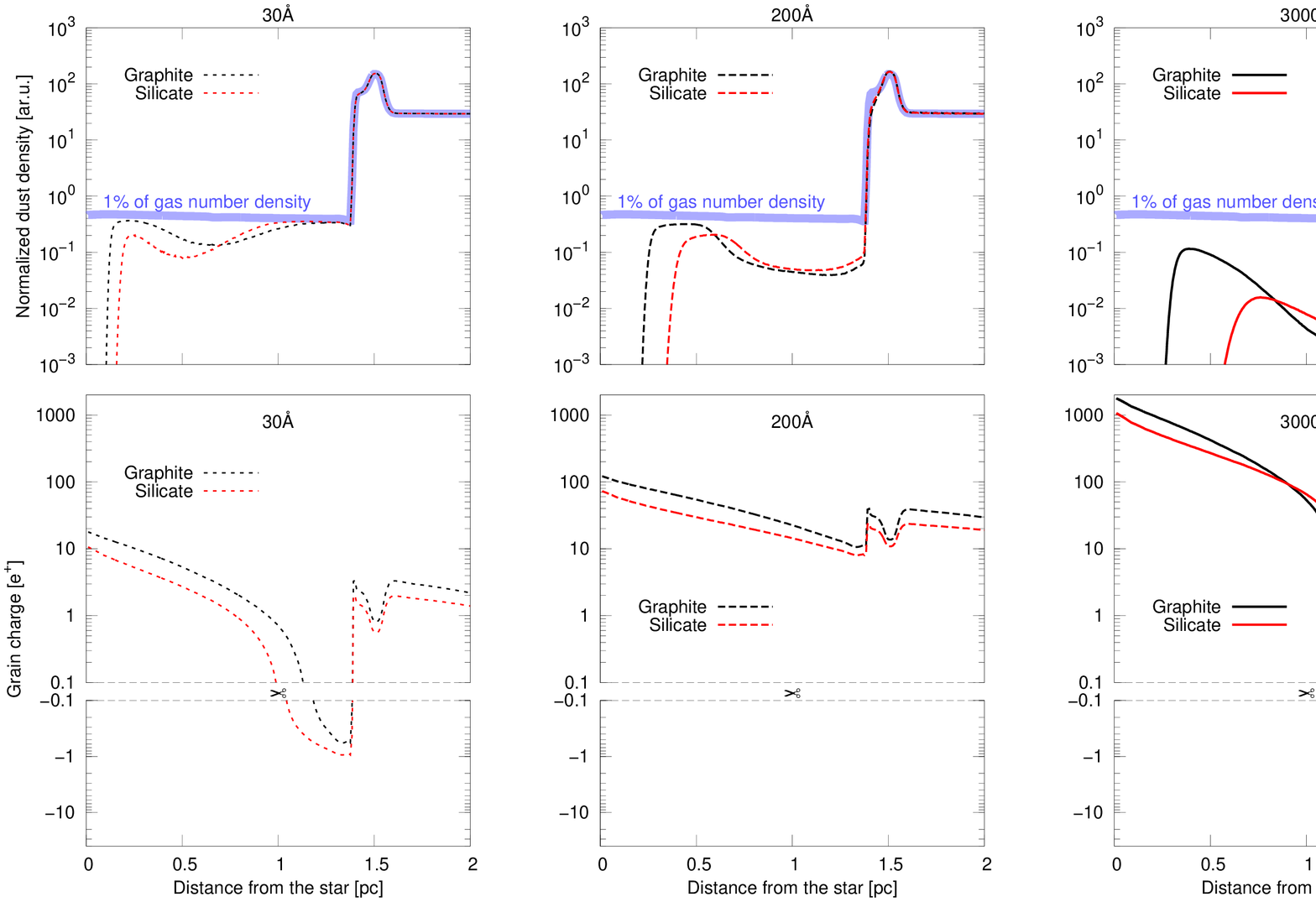}
\caption{Top panels: Density profiles of graphitic and silicate VSGs, ISGs and BGs normalized to 1~per~cent of the gas number density. Bottom panels: Their corresponding mean charges.}
\label{fig:chemcomp}
\end{figure*}

The impact of grain chemical composition on the dust drift is twofold. First, chemical composition affects grain charge both through different absorbing efficiencies and photoelectron work functions for graphites and silicates. Generally, as graphites have larger absorbing coefficients and lower photoelectron work function, they have larger charges and are more efficiently coupled to the gas. Second, this coupling is compensated somewhat by the stronger radiation pressure acting on the more absorbing graphitic grains. The overall effect of grain chemical composition on dust drift is shown in Fig.~\ref{fig:chemcomp} for the case when all the grains are assumed to have fluctuating charges. Graphitic grains tend to have more extended distributions and are better coupled to the gas in comparison with silicate grains, though the difference is not qualitative.

\section{Discussion}

The presented study is our next step \citep[following][]{paper_i} towards a consistent interpretation and explanation of observed IR intensity distributions in \hii\ regions around young stars. There is a general consensus that 8~\mkm\ emission is formed by PAHs, so that the ring-like 8~\mkm\ emission maps hint at the absence of PAHs in the interiors of \hii\ regions, where they are either destroyed or blown away by the stellar wind or by radiation pressure.

The presence of 24~\mkm\ emission in the interiors of \hii\ regions indicates that grains larger than PAHs may exist there, but the exact origin of the 24~\mkm\ emission is still a puzzle. According to \citet{paladini_12}, at $\lambda > 70$~\mkm\ IR spectra of 16 \hii\ regions up to 100~\mkm, obtained with {\it Herschel}, can be generated by big grains with \tdust\ about 20--30~K, while smaller grains with \tdust\ about 50--90~K are responsible for IR emission with $\lambda < 70$~\mkm.

Dynamical models of \hii\ regions show the importance of stellar winds, see e.g. reviews by \cite{2007dmsf.book..183A} and \cite{2007dmsf.book..103H}. However, it is not easy to explain how small grains may survive in a wind-blown bubble. This has led \cite{everett} to the conclusion that dust within an IR bubble may be constantly replenished from evaporating protoplanetary discs (dusty cloudlets) embedded in the \hii\ region. \cite{dustwave} related 24~\mkm\ emission to the material that is photo-evaporated from the inner bubble walls and then is shaped into a structure, which they coined a dust wave, by the radiation pressure of an exciting star. An important feature of the latter model is the pressure leakage from the bubble into the surrounding medium.

We explore whether dust distribution within an IR bubble can be explained without additional assumptions, like disc or inner wall evaporation, but simply due to dust evolution within a pressure-driven bubble. Specifically, in this paper we consider the effect of radiatively driven dust drift on the spatial distribution of different dust types. Our conclusion is that the combined action of radiation pressure and Coulomb drag on charged dust grains results in a significant stratification of dust of different sizes and thus can be a potential explanation for the observed IR morphology of \hii\ regions. This needs to be further supported by radiation transfer simulations, which will be presented in forthcoming papers.

Also, we have to keep in mind that there are other physical processes that are relevant in explaining the appearance of IR bubbles, for example, radiation pressure on the gas.  The stationary models of an \hii\ region  presented by \cite{Draine2011}, which account for the radiation pressure on the gas, demonstrate that density profiles inside the \hii\ region may have very different shapes depending on the model parameters. In particular, there are solutions with inner cavities that may be responsible for the deficit of dust in the interiors of \hii\ regions. The importance of radiation pressure on the gas and/or stellar wind is supported by the spatial correlation between 20~cm free--free emission and 24~\mkm\ dust emission towards some IR bubbles (e.g., N36 and N49; \cite{deharveng_10}).

Neither radiation pressure on the gas nor the stellar wind is taken into account in the presented model, and the density distribution of the ionized gas is nearly flat. Both factors, acting in concert with the radiation pressure on the dust, may be responsible for the absence of dust within the 24~\mkm\ ring. We plan to include the radiation pressure on the gas with the full dynamical coupling between the gas and dust it our future studies. But for the moment, our results indicate that a non-uniform dust distribution can appear even without these factors.

Another effect that may be important is the recycling/evolution of dust triggered by UV radiation and high temperature in \hii\ regions. According to \cite{Jones:2009}, the aromatisation and subsequent fragmentation of hydrogenated amorphous carbon grains can be a primary source of the aromatic emission band carriers. With this in mind, one may suggest that PAHs do not even need to be initially present in the molecular gas. The 8~\mkm\ emission can be generated by the transient layer of dust particles that are formed and quickly destroyed just behind the ionization front, rather than by PAHs that have been somehow removed from the ionized region. To check this idea, it is necessary to develop a dynamical model coupled with a model of dust microevolution.

\section{Conclusions}

In this paper we investigated dust dynamics in a pressure-driven expanding \hii\ region to understand the appearance of IR bubbles. As drag force and radiation pressure force strongly depend on dust size, their combined action leads to dust differentiation within an \hii\ region. We particularly focused on the effect of radiation pressure on charged grain motion and do not consider stellar wind and dust destruction. The grain charge is consistently evaluated to the compute aerodynamic drag force. Our conclusions can be summarized as follows:
\begin{itemize}
 \item Grain charge is of utter importance for dust velocity and the resultant density profiles \citep[see also][] {gail_79}). If all the grains are assumed to be neutral, the inner parts of the \hii\ region are almost devoid of dust. The grain charge increases dust friction with the gas and suppresses dust removal from the \hii\ region.
 \item The grain ensemble is characterized by the mean charge and charge dispersion. It is crucial to consider both parameters in the dynamics of small grains ($a<100$\,\AA). Coulomb drag for larger grains ($a>100$\,\AA) can be computed using their mean charge.
 \item The interplay between radiation pressure and drag generally leads to double-peak dust density profiles. The outer peak corresponds to the gas density maximum on the ionization and dissociation fronts. The inner peak arises due to the negative gradient of grain charge within the ionized region. Grains in the outer parts of the region have smaller charges and are less coupled to the gas. Sweeping of these grains leads to a density dip between the inner ionized region and the outer density maximum. The details of this process strongly depend on grain size.
 \item {\it PAHs}, having the largest surface-to-mass ratio, are well coupled to the gas whenever they have a non-zero charge. PAHs with zero charge are easily blown away from the star's vicinity. Near-zero charge fluctuations during the expansion lead to removal of PAHs from the inner part of the \hii\ region. So, in principle, the lack of 8~$\mu$m emission inside the \hii\ region can be explained not only by the destruction of PAHs but also by the radiative removal of PAHs if their charge fluctuates around zero.
 \item {\it Very small grains} acquire zero charge less frequently than PAHs and, consequently, are better coupled to the gas and do not escape ionized region easily. The density distribution of VSGs is the most uniform relative to other grain types. Stochastically heated VSGs may partially (along with ISGs) explain the double-peak morphology of IR bubbles at 24~$\mu$m. 
\item The intriguing feature of {\it intermediate-sized grains} ($a\approx200$\,\AA) is that their emission could contribute to both mid- and far-IR bands. The inner peak on the ISG density distribution is several orders of magnitude smaller than the outer peak, but the negative dust temperature gradient may lead to an inverse ratio of corresponding intensities at 24~$\mu$m. At the same time, the outer peak in the ISG distribution may be observable at 70--100~$\mu$m.
\item {\it Big grains} are subject to the strongest sweeping by the radiation pressure. Being able to contribute to far-IR, they are worthy candidates for the outer ring for emission at 70--100~$\mu$m.
\end{itemize}
The real dust size distribution in \hii\ regions is more complicated than described by the four dust types listed above. However, consideration of PAHs, VSGs, ISGs and BGs is necessary to understand the contribution of different subsamples of the grain size distribution in the dust emission spectrum.  While such problems are generally ill defined, additional information can be retrieved from the \hii\ region morphology accompanied by  grain dynamics modelling.
\section*{Acknowledgments}

We thank an anonymous referee for her/his stimulating report. This work is supported by the President of the Russian Federation grants (MK-2570.2014.2 and NSh-3620.2014.2), the Dynasty foundation and by RFBR grant 13-02-00642. We are grateful to L.~Deharveng, A.~Zavagno, W.~Henney, and M.~Murga for useful discussions and comments.

\newpage
\appendix
\section{Grain charge} \label{AppA}
The grain charge is primarily determined by photoelectric emission and accretion of ions and electrons. The probability $f(Z_{\rm gr})$ of finding a grain having charge $Z_{\rm gr}e$ can be determined from the equation of detailed balance~\citep{tielensbook}:
\begin{align}
 f(Z_{\rm gr})&\left[J_{\rm pe}(Z_{\rm gr})+J_{\rm ion}(Z_{\rm gr})\right]  = \nonumber \\
 &= f(Z_{\rm gr}+1)J_{\rm e}(Z_{\rm gr}+1),
\end{align}
where $J_{\rm pe}$~[\,s$^{-1}$] is a photoelectric emission rate, $J_{\rm e}$ and $J_{\rm ion}$~[\,s$^{-1}$] are the electron and ion accretion rates, respectively. Writing down the above equation for every $Z_{\rm gr}$, one finds an equation for $f(Z_{\rm gr})$:
 \begin{equation} 
  f(Z_{\rm gr})=f_0\prod\limits_{z=Z_0+1}^{Z_{\rm gr}}\frac{J_{\rm pe}(z-1)+J_{\rm ion}(z-1)}{J_{\rm e}(z)} \label{fZ1}
 \end{equation}
for $Z_{\rm gr}>Z_0$ and
\begin{equation}
 f(Z_{\rm gr})=f_0\prod\limits_{z=Z_{\rm gr}}^{Z_0-1}\frac{J_{\rm e}(z+1)}{J_{\rm pe}(z)+J_{\rm ion}(z)}  \label{fZ2}
\end{equation}
for $Z_{\rm gr}<Z_0$. Here $f_0\equiv f(Z_0)$. From the computational point of view, it is important that $Z_0$ should correspond to the maximum of $f(Z_{\rm gr})$, which can be estimated from the equation:
\begin{equation}
 J_{\rm pe}(Z_0)\approx J_{\rm e}(Z_0). \label{Zapp}
\end{equation}
Equations \eqref{fZ1}--\eqref{fZ2} are closed by
\begin{equation}
 \sum\limits_{z=-\infty}^{\infty} f(z) =1.
\end{equation}
Photon attenuation depth and electron mean free path are assumed to be equal to 100\AA and 10\AA, respectively.

The accretion rate of particles with number density $n$ and mass $m$ was adopted from~\citet{1987ApJ...320..803D}:
\begin{equation}
 J_{\rm acc}=n\sqrt{\frac{8kT_{\rm gas}}{\pi m}}\pi a_{\rm gr}^2 \tilde J (a_{\rm gr},T_{\rm gas},Z_{\rm gr}).
\end{equation}
Here $\tilde{J}$ accounts for Coulomb focusing. As sticking probabilities are not well known, we follow \citet{2001ApJS..134..263W} in assuming sticking probabilities for the electrons and ions to be equal to 0.5 and 1, respectively.

The photoemission rate depends on the mean intensity $J_{\nu}$~[\,erg/(cm$^2$ s Hz sr) ] and photo-ionization yield $Y$:
\begin{equation}
 J_{\rm pe}=\int \frac{4\pi J_{\nu}}{h\nu}C_{\rm abs}(\nu,a_{\rm gr}) Y(\nu,Z_{\rm gr},a_{\rm gr})\,d\nu
\end{equation}
 and is calculated according to \cite{2001ApJS..134..263W}. 

\label{lastpage}

\bibliographystyle{mn2e} 
\bibliography{paper_references}

\begin{thebibliography}{67}
\expandafter\ifx\csname natexlab\endcsname\relax\def\natexlab#1{#1}\fi

\bibitem[{{Aannestad} \& {Emery}(2001)}]{2001A&A...376.1040A}
{Aannestad} P.~A., {Emery} R.~J., 2001, \aap, 376, 1040

\bibitem[{{Aitken}, {Griffiths} \& {Jones}(1977){Aitken}, {Griffiths}, \&
  {Jones}}]{aitken}
{Aitken} D.~K., {Griffiths} J., {Jones} B., 1977, \mnras, 179, 179

\bibitem[{{Anderson} {et~al}\mbox{.}(2012){Anderson}, {Zavagno}, {Deharveng},
  {Abergel}, {Motte}, {Andr{\'e}}, {Bernard}, {Bontemps}, {Hennemann}, {Hill},
  {Rod{\'o}n}, {Roussel}, \& {Russeil}}]{Anderson_12}
{Anderson} L.~D. {et~al.}, 2012, \aap, 542, A10

\bibitem[{{Arthur}(2007)}]{2007dmsf.book..183A}
{Arthur} S.~J., 2007, {Wind-Blown Bubbles around Evolved Stars}, {Hartquist}
  T.~W., {Pittard} J.~M., {Falle} S.~A.~E.~G., eds., p. 183

\bibitem[{{Arthur} {et~al}\mbox{.}(2011){Arthur}, {Henney}, {Mellema}, {de
  Colle}, \& {V{\'a}zquez-Semadeni}}]{2011MNRAS.414.1747A}
{Arthur} S.~J., {Henney} W.~J., {Mellema} G., {de Colle} F.,
  {V{\'a}zquez-Semadeni} E., 2011, \mnras, 414, 1747

\bibitem[{{Arthur} {et~al}\mbox{.}(2004){Arthur}, {Kurtz}, {Franco}, \&
  {Albarr{\'a}n}}]{Arthuretal2004}
{Arthur} S.~J., {Kurtz} S.~E., {Franco} J., {Albarr{\'a}n} M.~Y., 2004, \apj,
  608, 282

\bibitem[{{Chini}, {Kruegel} \& {Wargau}(1987){Chini}, {Kruegel}, \&
  {Wargau}}]{chini_87}
{Chini} R., {Kruegel} E., {Wargau} W., 1987, \aap, 181, 378

\bibitem[{{Churchwell} {et~al}\mbox{.}(2006){Churchwell}, {Povich}, {Allen},
  {Taylor}, {Meade}, {Babler}, {Indebetouw}, {Watson}, {Whitney}, {Wolfire},
  {Bania}, {Benjamin}, {Clemens}, {Cohen}, {Cyganowski}, {Jackson},
  {Kobulnicky}, {Mathis}, {Mercer}, {Stolovy}, {Uzpen}, {Watson}, \&
  {Wolff}}]{churchwell_06}
{Churchwell} E. {et~al.}, 2006, \apj, 649, 759

\bibitem[{{Compi{\`e}gne} {et~al}\mbox{.}(2011){Compi{\`e}gne}, {Verstraete},
  {Jones}, {Bernard}, {Boulanger}, {Flagey}, {Le Bourlot}, {Paradis}, \&
  {Ysard}}]{DIMmodel}
{Compi{\`e}gne} M. {et~al.}, 2011, \aap, 525, A103

\bibitem[{{Crampton}(1971)}]{crampton_71}
{Crampton} D., 1971, \aj, 76, 260

\bibitem[{{Deharveng} {et~al}\mbox{.}(2010){Deharveng}, {Schuller}, {Anderson},
  {Zavagno}, {Wyrowski}, {Menten}, {Bronfman}, {Testi}, {Walmsley}, \&
  {Wienen}}]{deharveng_10}
{Deharveng} L. {et~al.}, 2010, \aap, 523, A6

\bibitem[{{Deharveng}, {Zavagno} \& {Caplan}(2005){Deharveng}, {Zavagno}, \&
  {Caplan}}]{dehaveng_05}
{Deharveng} L., {Zavagno} A., {Caplan} J., 2005, \aap, 433, 565

\bibitem[{{Deharveng} {et~al}\mbox{.}(2009){Deharveng}, {Zavagno}, {Schuller},
  {Caplan}, {Pomar{\`e}s}, \& {De Breuck}}]{deharveng_09}
{Deharveng} L., {Zavagno} A., {Schuller} F., {Caplan} J., {Pomar{\`e}s} M., {De
  Breuck} C., 2009, \aap, 496, 177

\bibitem[{{Diaz-Miller}, {Franco} \& {Shore}(1998){Diaz-Miller}, {Franco}, \&
  {Shore}}]{diaz_miller}
{Diaz-Miller} R.~I., {Franco} J., {Shore} S.~N., 1998, \apj, 501, 192

\bibitem[{{Draine}(2011{\natexlab{a}})}]{Draine2011}
{Draine} B.~T., 2011{\natexlab{a}}, \apj, 732, 100

\bibitem[{{Draine}(2011{\natexlab{b}})}]{drainebook}
{Draine} B.~T., 2011{\natexlab{b}}, {Physics of the Interstellar and
  Intergalactic Medium}. Princeton University Press, 2011.~ISBN:
  978-0-691-12214-4

\bibitem[{{Draine} \& {Li}(2007)}]{2007ApJ...657..810D}
{Draine} B.~T., {Li} A., 2007, \apj, 657, 810

\bibitem[{{Draine} \& {Salpeter}(1979)}]{DraineSalpeter1979}
{Draine} B.~T., {Salpeter} E.~E., 1979, \apj, 231, 77

\bibitem[{{Draine} \& {Sutin}(1987)}]{1987ApJ...320..803D}
{Draine} B.~T., {Sutin} B., 1987, \apj, 320, 803

\bibitem[{{Everett} \& {Churchwell}(2010)}]{everett}
{Everett} J.~E., {Churchwell} E., 2010, \apj, 713, 592

\bibitem[{{Ferland} {et~al}\mbox{.}(2013){Ferland}, {Porter}, {van Hoof},
  {Williams}, {Abel}, {Lykins}, {Shaw}, {Henney}, \& {Stancil}}]{cloudy_13}
{Ferland} G.~J. {et~al.}, 2013, Rev. Mex. Ast., 49, 137

\bibitem[{{Gail} \& {Sedlmayr}(1979)}]{gail_79}
{Gail} H.~P., {Sedlmayr} E., 1979, \aap, 77, 165

\bibitem[{{Gendelev} \& {Krumholz}(2012)}]{2012ApJ...745..158G}
{Gendelev} L., {Krumholz} M.~R., 2012, \apj, 745, 158

\bibitem[{{Georgelin} \& {Georgelin}(1970)}]{georgelin_70}
{Georgelin} Y.~P., {Georgelin} Y.~M., 1970, \aaps, 3, 1

\bibitem[{{Gillett} {et~al}\mbox{.}(1975){Gillett}, {Forrest}, {Merrill},
  {Soifer}, \& {Capps}}]{Gillett1975}
{Gillett} F.~C., {Forrest} W.~J., {Merrill} K.~M., {Soifer} B.~T., {Capps}
  R.~W., 1975, \apj, 200, 609

\bibitem[{{Harper} \& {Low}(1971)}]{hl71}
{Harper} D.~A., {Low} F.~J., 1971, \apjl, 165, L9

\bibitem[{{Harvey-Smith}, {Madsen} \& {Gaensler}(2011){Harvey-Smith}, {Madsen},
  \& {Gaensler}}]{2011ApJ...736...83H}
{Harvey-Smith} L., {Madsen} G.~J., {Gaensler} B.~M., 2011, \apj, 736, 83

\bibitem[{{Henney}(2007)}]{2007dmsf.book..103H}
{Henney} W.~J., 2007, {How to Move Ionized Gas: An Introduction to the Dynamics
  of HII Regions}, {Hartquist} T.~W., {Pittard} J.~M., {Falle} S.~A.~E.~G.,
  eds., p. 103

\bibitem[{{Hollenbach} \& {McKee}(1979)}]{hollenbach79}
{Hollenbach} D., {McKee} C.~F., 1979, \apjs, 41, 555

\bibitem[{{Hollenbach} \& {McKee}(1989)}]{hollenbach89}
{Hollenbach} D., {McKee} C.~F., 1989, \apj, 342, 306

\bibitem[{{Hosokawa} \& {Inutsuka}(2006)}]{hosokawa_06}
{Hosokawa} T., {Inutsuka} S.-i., 2006, \apj, 646, 240

\bibitem[{{Jones}(2009)}]{Jones:2009}
{Jones} A.~P., 2009, in Astronomical Society of the Pacific Conference Series,
  Vol. 414, Cosmic Dust - Near and Far, {Henning} T., {Gr{\"u}n} E.,
  {Steinacker} J., eds., p. 473

\bibitem[{{Jonkheid} {et~al}\mbox{.}(2004){Jonkheid}, {Faas}, {van Zadelhoff},
  \& {van Dishoeck}}]{jonkheid04}
{Jonkheid} B., {Faas} F.~G.~A., {van Zadelhoff} G.-J., {van Dishoeck} E.~F.,
  2004, \aap, 428, 511

\bibitem[{{Kirsanova}, {Wiebe} \& {Sobolev}(2009){Kirsanova}, {Wiebe}, \&
  {Sobolev}}]{paper_hiimodel}
{Kirsanova} M.~S., {Wiebe} D.~S., {Sobolev} A.~M., 2009, Astronomy Reports, 53,
  611

\bibitem[{{Krumholz} \& {Matzner}(2009)}]{KrumholzMatzner2009}
{Krumholz} M.~R., {Matzner} C.~D., 2009, \apj, 703, 1352

\bibitem[{{Krumholz}, {Stone} \& {Gardiner}(2007){Krumholz}, {Stone}, \&
  {Gardiner}}]{2007ApJ...671..518K}
{Krumholz} M.~R., {Stone} J.~M., {Gardiner} T.~A., 2007, \apj, 671, 518

\bibitem[{{Kurucz}(1979)}]{kurucz_79}
{Kurucz} R.~L., 1979, \apjs, 40, 1

\bibitem[{{Mackey} \& {Lim}(2011)}]{2011MNRAS.412.2079M}
{Mackey} J., {Lim} A.~J., 2011, \mnras, 412, 2079

\bibitem[{{Martins} {et~al}\mbox{.}(2010){Martins}, {Pomar{\`e}s}, {Deharveng},
  {Zavagno}, \& {Bouret}}]{martins_2010}
{Martins} F., {Pomar{\`e}s} M., {Deharveng} L., {Zavagno} A., {Bouret} J.~C.,
  2010, \aap, 510, A32

\bibitem[{{Mathews}(1965)}]{mathews_65}
{Mathews} W.~G., 1965, \apj, 142, 1120

\bibitem[{{Mathews}(1967)}]{mathews_67}
{Mathews} W.~G., 1967, \apj, 147, 965

\bibitem[{{Millar}, {Farquhar} \& {Willacy}(1997){Millar}, {Farquhar}, \&
  {Willacy}}]{umist_95}
{Millar} T.~J., {Farquhar} P.~R.~A., {Willacy} K., 1997, \aaps, 121, 139

\bibitem[{{Nakano} {et~al}\mbox{.}(1983){Nakano}, {Kogure}, {Sasaki}, {Mizuno},
  {Sakka}, \& {Wiramihardja}}]{Nakano1983}
{Nakano} M., {Kogure} T., {Sasaki} T., {Mizuno} S., {Sakka} K., {Wiramihardja}
  S.~D., 1983, Ap\&SS, 89, 407

\bibitem[{{Neufeld} \& {Kaufman}(1993)}]{neufeld93}
{Neufeld} D.~A., {Kaufman} M.~J., 1993, \apj, 418, 263

\bibitem[{{Ney} \& {Allen}(1969)}]{na69}
{Ney} E.~P., {Allen} D.~A., 1969, \apjl, 155, L193

\bibitem[{{Ochsendorf} {et~al}\mbox{.}(2014){Ochsendorf}, {Verdolini}, {Cox},
  {Bern{\'e}}, {Kaper}, \& {Tielens}}]{dustwave}
{Ochsendorf} B.~B., {Verdolini} S., {Cox} N.~L.~J., {Bern{\'e}} O., {Kaper} L.,
  {Tielens} A.~G.~G.~M., 2014, \aap, 566, A75

\bibitem[{{O'Dell} \& {Hubbard}(1965)}]{odell_orion}
{O'Dell} C.~R., {Hubbard} W.~B., 1965, \apj, 142, 591

\bibitem[{{O'Dell}, {Hubbard} \& {Peimbert}(1966){O'Dell}, {Hubbard}, \&
  {Peimbert}}]{odell_others}
{O'Dell} C.~R., {Hubbard} W.~B., {Peimbert} M., 1966, \apj, 143, 743

\bibitem[{{Paladini} {et~al}\mbox{.}(2012){Paladini}, {Umana}, {Veneziani},
  {Noriega-Crespo}, {Anderson}, {Piacentini}, {Pinheiro Gon{\c c}alves},
  {Paradis}, {Tibbs}, {Bernard}, \& {Natoli}}]{paladini_12}
{Paladini} R. {et~al.}, 2012, \apj, 760, 149

\bibitem[{{Panagia}(1974)}]{panagia}
{Panagia} N., 1974, \apj, 192, 221

\bibitem[{{Pavlyuchenkov}, {Kirsanova} \& {Wiebe}(2013){Pavlyuchenkov},
  {Kirsanova}, \& {Wiebe}}]{paper_i}
{Pavlyuchenkov} Y.~N., {Kirsanova} M.~S., {Wiebe} D.~S., 2013, Astronomy
  Reports, 57, 573

\bibitem[{{Petrosian}, {Silk} \& {Field}(1972){Petrosian}, {Silk}, \&
  {Field}}]{1972ApJ...177L..69P}
{Petrosian} V., {Silk} J., {Field} G.~B., 1972, \apjl, 177, L69

\bibitem[{{Russeil}(2003)}]{russeil_03}
{Russeil} D., 2003, \aap, 397, 133

\bibitem[{{Simpson} {et~al}\mbox{.}(2012){Simpson}, {Povich}, {Kendrew},
  {Lintott}, {Bressert}, {Arvidsson}, {Cyganowski}, {Maddison}, {Schawinski},
  {Sherman}, {Smith}, \& {Wolf-Chase}}]{simpson_12}
{Simpson} R.~J. {et~al.}, 2012, \mnras, 424, 2442

\bibitem[{{Spitzer}(1978)}]{spitzerbook}
{Spitzer} L., 1978, {Physical processes in the interstellar medium}. New York
  Wiley-Interscience, 1978.~333 p.

\bibitem[{{Stein} \& {Gillett}(1969)}]{sg69}
{Stein} W.~A., {Gillett} F.~C., 1969, \apjl, 155, L197

\bibitem[{{Stone} \& {Norman}(1992)}]{stone_92}
{Stone} J.~M., {Norman} M.~L., 1992, \apjs, 80, 753

\bibitem[{{Tenorio-Tagle}(1974)}]{tt74}
{Tenorio-Tagle} G., 1974, Ap\&SS, 26, 111

\bibitem[{{Tielens}(2005)}]{tielensbook}
{Tielens} A.~G.~G.~M., 2005, {The Physics and Chemistry of the Interstellar
  Medium, Cambridge, UK: Cambridge University Press}. Cambridge University
  Press

\bibitem[{{Tielens} \& {Hollenbach}(1985)}]{tielens_85}
{Tielens} A.~G.~G.~M., {Hollenbach} D., 1985, \apj, 291, 722

\bibitem[{{van Dishoeck}(1988)}]{vdh_88}
{van Dishoeck} E.~F., 1988, in Astrophysics and Space Science Library, Vol.
  146, Rate Coefficients in Astrochemistry, {Millar} T.~J., {Williams} D.~A.,
  eds., pp. 49--72

\bibitem[{{van Dishoeck}, {Jonkheid} \& {van Hemert}(2006){van Dishoeck},
  {Jonkheid}, \& {van Hemert}}]{faraday_06}
{van Dishoeck} E.~F., {Jonkheid} B., {van Hemert} M.~C., 2006, Faraday
  Discussions, 133, 231

\bibitem[{{Watson} {et~al}\mbox{.}(2008){Watson}, {Povich}, {Churchwell},
  {Babler}, {Chunev}, {Hoare}, {Indebetouw}, {Meade}, {Robitaille}, \&
  {Whitney}}]{watson_08}
{Watson} C. {et~al.}, 2008, \apj, 681, 1341

\bibitem[{{Weingartner} \& {Draine}(2001)}]{2001ApJS..134..263W}
{Weingartner} J.~C., {Draine} B.~T., 2001, \apjs, 134, 263

\bibitem[{{Wright}(1973)}]{wright}
{Wright} E.~L., 1973, \apj, 185, 569

\bibitem[{{Zavagno} {et~al}\mbox{.}(2007){Zavagno}, {Pomar{\`e}s}, {Deharveng},
  {Hosokawa}, {Russeil}, \& {Caplan}}]{zavagno_07}
{Zavagno} A., {Pomar{\`e}s} M., {Deharveng} L., {Hosokawa} T., {Russeil} D.,
  {Caplan} J., 2007, \aap, 472, 835

\bibitem[{{Zavagno} {et~al}\mbox{.}(2010){Zavagno}, {Russeil}, {Motte},
  {Anderson}, {Deharveng}, {Rod{\'o}n}, {Bontemps}, {Abergel}, {Baluteau},
  {Sauvage}, {Andr{\'e}}, {Hill}, \& {White}}]{zavagno_10}
{Zavagno} A. {et~al.}, 2010, \aap, 518, L81

\end{thebibliography}

\end{document}